\documentclass[preprint,review,12pt]{elsarticle}
\usepackage{amssymb}
\usepackage{bm}
\usepackage{ulem}
\usepackage{xcolor}
\usepackage{lineno}

\begin{document}

\begin{frontmatter}

\title{Heat balance model for long-term prediction of the thermal regime of a subway tunnel and surrounding soil}
\author[a]{G.P. Vasillyev}
\author[b]{N.V. Peskov\corref{*}} 
\ead{peskov@cs.msu.ru}
\cortext[*]{Corresponding author}
\author[b]{T.M. Lysak} 
\address[a]{OAO “INSOLAR-INVEST”, Moscow, Russian Federation.}
\address[b]{Faculty of Computational Mathematics and Cybernetics, \\Lomonosov Moscow State University,  Moscow, Russian Federation.}

\begin{abstract}

For a long-term forecast of the average air temperature in a metro tunnel and surrounding soil, heat balance conditions are modeled in a system that includes two parallel tunnels, a near-tunnels underground space, the earth's surface and atmospheric air with seasonal temperature variation. It is assumed that the air in the tunnels is well mixed, and the thermal effect of ventilation is taken into account by replacing the air in the tunnels with atmospheric air. The system of heat balance equations is solved numerically by the finite element method. A numerical analysis of the influences of the model parameters on the average temperature in the tunnels and surrounding soil is carried out. The dimensions of the soil region with an elevated temperature, located between the tunnels, are estimated. 

\end{abstract}


\end{frontmatter}
%\linenumbers

\section{Introduction}

Among the public transport systems in urban areas, the subway is one of the most advanced due to its high throughput, low operating costs and the area of premises for maintenance  \cite{1}. When designing and operating the subway system, special attention is paid to maintaining the temperature and humidity in the tunnel within certain limits. These limits are determined by the conditions of human comfort and the technological requirements of the installed equipment \cite{2}. Along with defining principles and approaches to the design and management of the subway, presented in [2] in the USA,  a complex of computer programs, known as Subway Environment Simulation (SES),  was developed  by 1975. The fourth version of SES  was published in 1997 \cite{3}, and it is still the most common simulation software. SES is intended mainly for short-term modeling and control of environmental parameters. However, it provides the possibility of a long-term forecast of the thermal regime of the tunnel, though the long-term prediction within SES is based on a quasi-stationary solution of the 1D (radial) model of a homogeneous underground space. Therefore, SES cannot model the transient and it does not take into account the heat flow through the earth's surface, which can be significant for shallow tunnels.
Another application for simulation of environmental conditions and airflows in a subway system is IDA Tunnel developed by EQUA in 1995 \cite{4}. IDA Tunnel  is based on  SES modeling approach, and it also allows only 1D simulation of large scales problems. 

Apparently, the main difficulty in calculating the temperature in the tunnel is to adequately take into account  the heterogeneity of the soil and the heat transfer by turbulent air flows in the tunnel, caused by the movement of trains and ventilation. In some studies, the tunnel temperature was calculated analytically with some significant simplifications of the problem. For example, in \cite{5}, the temperature of the air in the tunnel is calculated assuming the constant speed of the air flow and the quasi-stationary heat exchange between the tunnel and the surrounding homogeneous soil. An analytical solution was found in \cite{6}  for transient heat conduction in a multilayer annulus. Analytical methods provide an exact solution to the problems, but the simplifications adopted in this case significantly limit the practical application of results.

There are many studies devoted to the numerical modeling of the heat and moisture production and transfer in tunnels. We mention some of these works. In \cite{7}, the finite element method was applied in a 3D model to determine the air temperature in a mountain tunnel and the temperature of  the rock, surrounding the tunnel, with taking into account freezing-thawing of ground moisture. In this case, the air flow in the tunnel was assumed to be laminar, and the heat source or absorbers in the tunnel were not taken into account. The problem of heat transfer in the ventilation tunnel connecting the surface of the earth and the underground room, and the surrounding soil was solved in \cite{8} using the finite difference method. The finite difference method was also used in \cite{9} to simulate the thermal regime of a mountain transport tunnel, taking into account natural and mechanical ventilation and wind caused by the movement of trains. In \cite{10}, the Green function method is proposed to solve the problem of thermal conductivity in the soil mass, and the Green function is calculated by the finite element method since it is determined under fixed boundary conditions. The evolution of air temperature in the tunnel is described by the finite-difference transfer equation, taking into account the heat production in the tunnel and the heat exchange with the soil. 

Recently, the program STESS \cite{11} was developed, the Chinese version of the program SES, which improves the calculation of non-stationary aerodynamic and thermal processes. In addition to theoretical studies, experimental studies have also been made. In particular, the influence of various factors on the temperature in the tunnel was investigated in \cite{12}. And also, new models and algorithms for optimal metro control are being developed \cite{13,14}.
Of the few experimental works devoted to the study of the thermal regime in tunnels and at metro stations, we note the work \cite{15} in which the thermal reaction of the soil surrounding the tunnel to periodic changes in the air temperature in the tunnel was experimentally studied.

In this paper, we propose a new model for the long-term forecast of the thermal regime of the tunnel and surrounding soil, taking into account heat generation and mechanical ventilation of the tunnel. It seems that a detailed long-term prediction of the temperature fields in the tunnel and in the surrounding soil is hardly possible. However, it is possible to predict the evolution of the average temperature, calculated  over the tunnel  volume, using average (effective) values of parameters.  

Given that most underground transportion systems use two parallel tunnels  for the trains traffic between the stations, we  consider a  system of two parallel tunnels in general position, located at a shallow depth. The shallow depth  of the tunnels requires taking into account the heat exchange between the atmospheric air and the ground surface.  Strong anisotropy of the temperature field and the  influence of the earth's surface distinguish our approach from most papers (see, for example, \cite{7}-\cite{9}, \cite{15}), in which a deep tunnel is considered in cylindrical coordanates statement and  the influence of the earth's surface temperature is neglected. 

The initial boundary value problem for the heat equation in the soil surrounding the tunnel is solved by the finite element method, which allows one to take into account the heterogeneity of the soil composition. For the average air temperature in the tunnel, a simple heat balance equation is written. To demonstrate the capabilities of the heat balance model, we simulate the thermal regime of a typical section of a transport tunnel, varying the values of the model parameters in fairly wide intervals that overlap their real values. Based on the simulation results, we estimate the volume of soil with elevated remperature (``the heat reservoir''  \cite{15})  and show that the permanent temperature core of the heat reservoir is located between the tunnels.

\section{Formulation of the problem}

Consider a subway system with two parallel circular single-track tunnels of radius $R_t$, the distance between the centers of which is $D_t$ and which are at a depth of $H_t$ from the earth's surface. Assuming for simplicity that the structure and composition of the soil around each tunnel is approximately the same, the thermal field in the vicinity of the tunnels can be considered symmetrical with respect to the vertical plane passing through the middle of the distance between the tunnels. 

\begin{figure}[hp]
	\centering
	\includegraphics{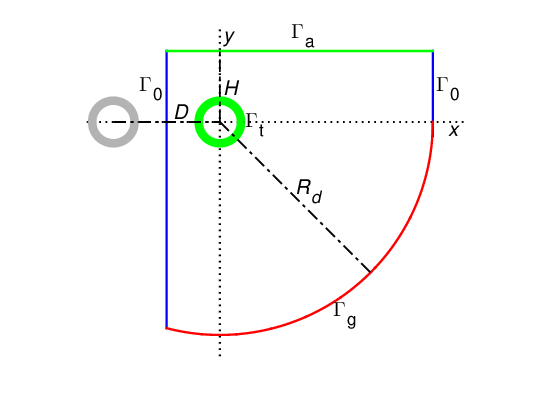}
	\caption{Computational domain}
	\label{fig_dom}
\end{figure}

By virtue of this assumption, all calculations will be performed for only one tunnel. Thus, the computational domain $G$ shown in Figure \ref{fig_dom} is a vertical soil layer surrounding one of the two tunnels and bounded by four boundary segments. The boundary segments are as follows: $\Gamma_t $ is the wall of the tunnel, $\Gamma_a$ is the horizontal surface of the earth, $\Gamma_g$ is the circular segment located at a distance $R_d$ from the center of the tunnel in the depth of the soil, and $\Gamma_0$ consists of two vertical straight segments connecting the ends of $\Gamma_g$ with the earth's surface. The presence of these vertical segments in the boundary of the domain is explained by the fact that we want to preserve the surface section $\Gamma_a$ in the model at any depth of the tunnel. Obviously, the domain can easily be generalized to an asymmetric case or to a double-track tunnel.

Further, we suppose the homogeneity of the characteristics of the tunnel and the surrounding soil along the  length of the tunnel, which means that the  temperature field does not dependent on  the third spatial coordinate. Thus, we reduce the three-dimensional  problem in space to a two-dimentional problem, assuming that the temperature depends only on the two spatial coordinates in the plane perpendicular to the tunnel. To determine the temperature field $T(x,y,t)$ ($x$, $y$ are the Cartesian coordinates in a vertical plane perpendicular to the tunnel, and $t$ is the time) in the outlined domain $G$ one has to solve the initial boundary value problem, that is composed from the following components. 

The heat transfer equation inside $G$
\begin{equation}
\label{heq}
C_g\frac{\partial T}{\partial t} = \nabla\cdot(\lambda_g\nabla T),
\end{equation}
where $C_g$ is the volumetric heat capacity and $\lambda_g$ is the heat conductivity coefficient of the soil. 

The boundary conditions
\begin{eqnarray}
\label{g0}
&& (\bm n \cdot \nabla T)\bigg{|} _{\Gamma_0} = 0,  \\
\label{gt} 
&& (\bm n\cdot\lambda_g\nabla T)\bigg{|} _{\Gamma_t} = \alpha_t(T_t-T), \\
\label{ga} 
&& (\bm n\cdot \lambda_g\nabla T)\bigg{|} _{\Gamma_a} = \alpha_a(T_a-T), \\
\label{gg}
&& T\bigg{|} _{\Gamma_g} = T_g;
\end{eqnarray}
where $\bm n$ is the external normal to the domain boundary. $\alpha_t$ and $\alpha_a$ are the heat exchange coefficients of tunnel air -- ground and ambient (atmospheric) air -- ground, respectively. $T_t(t)$ and $T_a(t)$ are the temperature of tunnel air and ambient air, respectively. $T_g$ is the constant temperature at the bottom of domain $G$.

Thus, the model assumes convective heat transfer on the tunnel wall $\Gamma_t$ and on the surface of the earth $\Gamma_a$, a constant temperature in the depth of the soil $\Gamma_g$, and zero heat flux through the boundary sections $\Gamma_0$. The distance $R_d$ should be taken large enough so that the temperature distortion near the tunnel due to temperature $T_g$ is small enough. Zero flux through the section of $\Gamma_0$ located between the tunnels follows from the symmetry of the temperature field, and the flux through the other section of $\Gamma_0$ has little effect on the temperature near the tunnel and is putted to zero for simplicity.

The initial temperature $T^0$ is set for the soil in the domain $G$
\begin{equation}
\label{ig}
T\bigg{|} _{t=0}=T^0(x,y), (x,y)\in G.
\end{equation}

The temperature of the ambient air in this problem is represented by a given function of time, $T_a(t)$, which can be obtained, in particular, from the weather archive for a given location. Thermophysical soil parameters are also input data for the problem. In contrast, the average temperature of the air in the tunnel, $T_t(t)$, is an unknown function of time, which needs to be determined as a result of solving the problem. 

We believe that three factors play a major role in assessing long-term trends in average tunnel air temperature: heat generation by various sources in the tunnel, ventilation using ambient air, and heat transfer between the air in the tunnel and the surrounding soil. These factors are described by the following quantities.

Heat production in the tunnel will be described by a time-dependent function $q(t)$, which is the average power of the heat sources over daily working time of the subway per 1 m of the tunnel length. The greatest amount of heat in the tunnel is generated during acceleration and braking of trains \cite{16}. Therefore, for long-term forecasting, the planned changes in traffic intensity can be taken into account when developing the function $q(t)$. 

Accurate modeling of airflows in a tunnel, generated by  ventilation, train movements and other causes,  is a very difficult task. However, we assume that for long-term assessment of average temperature it is sufficient to estimate the gross effect of ventilation on air temperature. We imagine this gross effect as a replacement per unit time of a certain volume of air in a tunnel with temperature $T_t$ by the same volume of outdoor air with temperature $T_a$. This process can be characterized by function $v(t)=\Delta V(t)/V_0$, where $V_0$ is the volume of the tunnel segment and $\Delta V$ is the volume of the air replaced by ventilation per unit time. 

Using these quantities, the heat balance equation for the air in the tunnel can be written as follows
\begin{equation}
\label{het}
C_a\frac{dT_t}{dt}=\frac{q}{S_t}+C_av(T_a-T_t) + \frac{2\alpha_t}{R_t}(\widetilde{T_s}-T_t),
\end{equation}
where $C_a$ is the volumetric (isobaric) heat capacity of air, $S_t$ is the tunnel cross-sectional area. The last term in the formula (\ref{het}) corresponds to the heat flux through the wall of the tunnel, which has the shape of a circular cylinder of radius $R_t$. The average temperature of the adjacent  soil in contact with the wall of the tunnel, $\widetilde{T_s}$ , is defined as follows,
\begin{equation}
\label{tav}
\widetilde{T_s}= \frac{1}{2\pi}\int_0^{2\pi}{T(\phi)\bigg{|} _{\Gamma_t}\,d\phi}.
\end{equation} 
The initial condition for Eq. (7) can be written in the form
\begin{equation}
\label{it}
T_t(0) = T_t^0.
\end{equation}

Equations (\ref{heq}), (\ref{het}) with initial conditions (\ref{ig}), (\ref{it}) and boundary conditions (\ref{g0})-(\ref{gg}) determine the evolution of the temperature field $T(x,y,t)$  of the surrounding soil and the average temperature  $T_t(t)$ in the tunnel.

\section{Numerical solution}

In order to solve system (\ref{heq})-(\ref{it}) numerically, it must be discretized. To discretize the heat equation (\ref{heq}), we use the finite element method (FEM). FEM is widely used in scientific and engineering calculations, and there is an extensive literature on the basics and applications of this method. For our purpose, it is sufficient to use FEM in its simplest form. Therefore, here we can outline the main points of the application of the method without referring to the special literature.

In this work, we consider the temperature $T_g$ at the boundary section$\Gamma_g$ to be constant. Therefore, it is convenient to make a replacement 
\[T(t,x,u) = T_g + U(t,x,y), \; T_t(t) = Tg + U_t(t).\] 
In this case, the equations and boundary conditions will retain their form, and the boundary condition (\ref{gg}) will become zero
\begin{equation}
\label{gg0}
U\bigg{|} _{\Gamma_g} = 0;
\end{equation}

\begin{figure}[hp]
	\centering
	\includegraphics{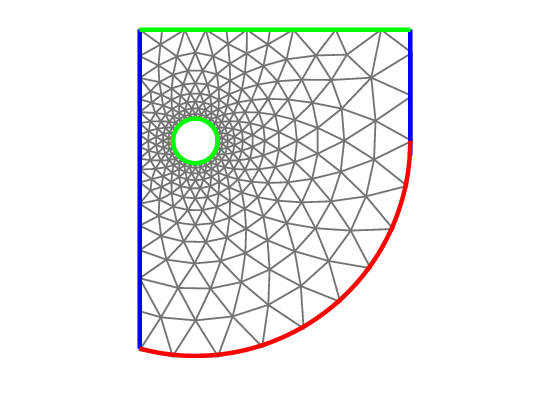}
	\caption{An example of triangular mesh in $G$.}
	\label{fig_grid}
\end{figure}

Discretization of the equation (\ref{heq}) begins with triangulation of the domain
$G$.  As an illustration, Figure \ref{fig_grid} shows an example of triangulation. In this example, the triangular mesh starts at the tunnel surface (circle $\Gamma_t$) and then continues to the boundaries of the domain with gradually increasing cell sizes. For clarity, the figure shows a mesh with 30 nodes on the circle $\Gamma_t$. In the calculations below, a finer mesh with 90 nodes on $\Gamma_t$ was used.

The nodes of the triangular mesh are denoted by $P_i$, $i=1,2,\dots,N_P$; $O_i$ is the neighborhood of the node $P_i$, i.e. the union of all triangles in which point $P_i$ is the vertex. For each node $P_i$, we define a linear finite element - a piecewise linear continuous function $\psi_i(x,y)$ such that it is linear in each triangle of the grid, is equal to 1 at point $P_i$, and is equal to zero outside $O_i$.  The function $\psi_i$ can be imagined as a pyramid with base $O_i$, triangular lateral faces and the top at point $P_i$.

Multiplying the equation (\ref{heq}) by $\psi_i$ and integrating over the domain $G$, we obtain the equality 
\[\int_G{\psi_iC_g\frac{\partial U}{\partial t}\,dxdy} = 
\int_G{\psi_i\nabla\cdot(\lambda_g\nabla U)\,dxdy}.\]
The integral on the right-hand side of this equality is transformed using Green's formula:
\begin{equation}
\label{wheq}
\int_G{\psi_iC_g\frac{\partial U}{\partial t}\,dxdy} = 
-\int_G{\nabla\psi_i\cdot(\lambda_g\nabla U)\,dxdy} +
\int_\Gamma{\psi_i(\bm n \cdot \lambda_g\nabla U)\,d\gamma} .
\end{equation}
Here the last integral is taken over the domain boundary $\Gamma$. Note that the flow $\bm n \cdot \lambda_g\nabla U$ at the border sections $\Gamma_0$, $\Gamma_t$ and $\Gamma_a$ is determined by the boundary conditions (\ref{g0})-(\ref{ga}).

The finite element approximation $U^{(f)}$ of function $U$ is defined as
\begin{equation}
\label{fea}
U(t,x,y) \approx U^{(f)}(t,x,y) = \sum_{j=1}^{N_p}{U_j(t)\psi_j(x,y)}.
\end{equation}
 The functions $U$ and $U^{(f)}$ coincide at the nodes of the grid, and inside each triangle the function $U^{(f)}$ is a linear interpolation of the values of the function $U$ at the vertices of the triangle. Replacing $U$ in the equation (\ref{wheq}) by $U^{(f)}$, we obtain a system of ordinary differential equations for $U_i(t)$, $i = 1,2,\dots,N_e$. The number of equations $N_e$ is equal to the number of grid nodes $N_p$ minus the number of nodes lying on the section of the boundary $\Gamma_g$, on which the values of $U_i$ are specified by the boundary condition
(\ref{gg0}).

In matrix form, the ODE system can be written as
\begin{equation}
\label{deq}
\mathbf M \frac{d\bm U}{dt} = \mathbf L \bm U + \bm F,
\end{equation}
where $\bm U(t) = (U_1(t), U_2(t),\dots, U_{N_e}(t))^{\mathrm T}$ is the column-vector of unknowns. Elements of $N_e\times N_e$ matrices $\mathbf M$ and $\mathbf L$ are independent of time and calculated by the formulas
\begin{equation}
\label{ml}
M_{ij} = \int_{O_i\cap O_j}{\psi_iC_g\psi_j\,dxdy}.
\end{equation}
\begin{equation}
\label{ll}
L_{ij} = \int_{O_i\cap O_j}{\nabla\psi_i\cdot\lambda_g\nabla\psi_j\,dxdy} + b_{ij},
\end{equation}
where $b_{ij}$ denotes the contribution of the boundary integral to the matrix element $L_{ij}$. $b_{ij}$ can be nonzero only if the nodes $P_i$ and $P_j$ lie on the boundary $\Gamma_a$ or $\Gamma_t$ and are calculated using the boundary conditions. Also, the component $F_i$ of the vector $\bm F$ 
can be nonzero only if $P_i$ lies on $\Gamma_a$ or $\Gamma_t$ and is calculated from the integral over the corresponding segment of the boundary using the boundary conditions (\ref {g0}) - (\ref {ga}), (\ref {gg0}).

Using the approximation (\ref{fea}), we transform the integral in (\ref{tav}) in a finite sum over the grid points belonging to $\Gamma_t$ and rewrite Eq. (\ref{het}) as follows
\begin{equation}
\label{det}
C_a\frac{dU_t}{dt} = \frac{q}{S_t}+C_av(T_a-T_g -U_t) + \frac{2\alpha_t}{R_t}\left( 
\frac{\Delta\phi}{2\pi}\sum_{p_j\in\Gamma_t}{U_j}-U_t\right).
\end{equation}
We assume that the grid points are uniformly distributed over the  circle $\Gamma_t$ and $\Delta\phi$ is the central angle between adjacent points.

Combining equations (\ref{deq}) and (\ref{det}), we obtain a closed system of $N_e+1$ linear differential equations. Our computational experience says that for typical values of the parameters, the system (\ref{deq}), (\ref{det}) has moderate stiffness and standard numerical methods can be used to solve it. In particular, the results presented below were obtained using the Crank-Nicholson scheme \cite{17} with a constant time step.

\section{Model input}

\subsection{Model parameters}

The model parameters are divided into geometrical parameters that determine the location and dimensions of tunnels, and thermophysical parameters of soil and air.  For the geometric parameters, we took values close to the typical values for the metro in Russia. Thermophysical parameters of the soil, heat capacity and thermal conductivity coefficient are considered constant, i.e. the soil is assumed to be homogeneous. Note that in FEM it is easy to take into account soil heterogeneity using coordinates-dependent values in the formulas (\ref{ml}). Constant values do not reduce the amount of calculations and do not simplify formulas, but they make it easier to interpret the results and detect errors in program codes. For brevity, we use the volumetric (isobaric) heat capacity of soil and air in our calculations. The values of geometrical and thermophysical parameters did not change from calculation to calculation and are presented in Table 1. The noticeable difference in the values of the heat exchange coefficients $\alpha_a$ and $\alpha_t$ is explained by the fact that when evaluating the value of $\alpha_t$, we tried to take into account the thermal resistance of the tunnel wall.

\begin{table}[h]
\caption{Non-variable model parameters}
\centering
\begin{tabular}{l c c c}
\hline
 Parameter & Value & Dimension \\
\hline
$C_g$ -- soil heat capacity & 2800 & kJ/(m$^3\cdot$K) \\
$\lambda_g$ -- soil heat conductivity & 1.0 & W/(m$\cdot$K) \\
$\alpha_a$ -- exchange soil-atmosphere & 20 & W/(m$^2\cdot$K) \\
$T_g$ -- soil temperature on $\Gamma_g$ & 10 & $^\circ$C \\
$\alpha_t$ -- exchange soil-tunnel & 5 & W/(m$^2\cdot$K) \\
$C_a$ -- air heat capacity & 1.21 & kJ/(m$^3\cdot$K) \\
$R_t$ -- tunnel radius & 3.0 & m \\
$H_t$ -- tunnel depth & 15 & m \\
$D_t$ -- distance between tunnels & 15 & m \\
$R_d$ -- radius of deep circle & 30 & m \\
\hline
\end{tabular}
\end{table}

In the numerical examples, the model variables are the parameters related to heat generation and tunnel ventilation. They are characterized by the total power of the various heat sources \cite{14}, $q(t)$, and the rate of replacement of air in the tunnel by ambient air, $v(t)$. These parameters are the control parameters of the model. Typically, a subway tunnel is operated for a certain period of time during the day. We assume that this period is 19 hours. Since the average daily temperature is used for calculations, the specific hours of operation of the tunnel do not matter. We believe that during the first 19 hours of each day, the parameters $q$ and $v$ have given constant values, and during the last 5 hours their values are zero.

\subsection{Ambien temperaturer, $T_a$}
\begin{figure}[hp]
	\centering
	\includegraphics{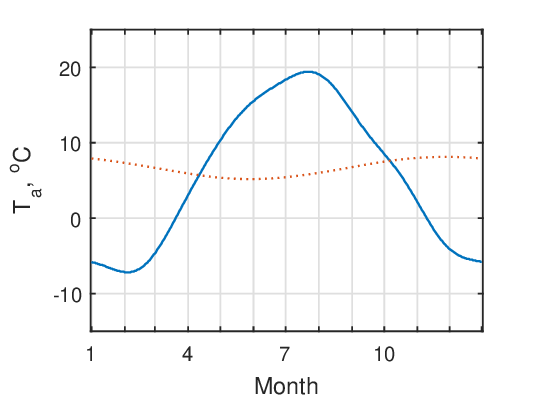}
	\caption{The average daily air temperature in Moscow for 1981-2010 years. Minimum: -7.2 -- 03.02, maximum: +19.4 -- 22.07. (http://meteoinfo.ru). Starting date is January 1. The dotted line is the ``natural'' soil temperature at a depth of 10 m ($x=0, y=5$). Vertical lines separate months.}
	\label{fig_TA}
\end{figure}

As the temperature of ambient air, $T_a(t)$, we used the average daily temperature in the city of Moscow for 1981-2010, presented by the Hydrometeorological Center of Russia (http://meteoinfo.ru). A graph of this temperature is shown in Figure \ref{fig_TA}.

\subsection{Initial conditions}

Although the problem solution quickly enough, for the first few years, "forgets" the initial temperature distribution (\ref{ig}), (\ref{it}), it is advisable to set the initial distribution as close as possible to the natural temperature distribution in the soil, taking into account seasonal variability, to avoid unnecessary artificial distortion of the solution at the beginning of the calculation.To establish the “natural” initial conditions, we solved the system (\ref{deq}), (\ref{det}) with the homogeneous initial temperature, $U (x, y, 0) = U_t (0) = 0$ and an “inactive” tunnel with $ q (t ) = v (t) = 0 $, for a sufficiently long time, until the annual temperature change becomes near periodic. We take the final temperature distribution for a specific calendar date (January 1, in  our examples) as a "natural" distribution for modeling thermal conditions of the active tunnel.

\begin{figure}[hp]
	\centering
	\includegraphics{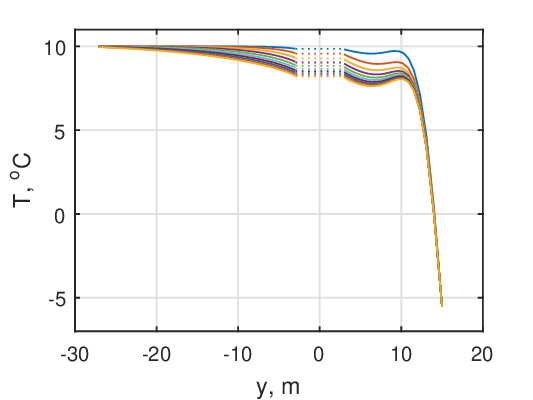}
	\caption{Soil temperature along the $y$ axis (see Fig. 1) on January 1st in 1st, 2nd, 3rd, ..., 10th years (from top to bottom) after starting counting with constant temperature 10$^\circ$C. The tunnel temperature is shown with a dotted line.}
	\label{fig_T0}
\end{figure}

Figure \ref{fig_T0} illustrates the calculation of the ``natural'' soil temperature. The figure shows the soil temperature along the $y$-axis (the vertical axis passing through the center of the tunnel) on January 1st, 1st, 2nd and subsequent years. At the beginning of the calculation (January 1 of the 0th year), the temperature was constant everywhere and equal to 10$^\circ$C. Over time, the temperature difference in successive years decreases. The temperature distribution converges to  ``natural''. We stopped the calculation when the maximum temperature difference became less than 0.1$^\circ$C. This happened in the 10th year.

Seasonal temperature variations are also observed in the near-surface soil. However, their amplitude decreases rapidly with increasing depth, and the ``phase shift'' between temperature variations in the soil and in the ambient air increases. In Figure \ref{fig_TA}, the dotted line shows a graph of "natural" soil temperature sesonal variations at a depth of 10m.

It should be noted that in all calculations with constant or cyclical parameters, a more or less prolonged transition period is observed, during which the temperature comes into dynamic equilibrium with the soil and the ambient air. We will assume that the transition period ends when the maximum temperature difference in successive years becomes less than 0.1$^\circ$C. All further results were obtained for the equilibrium state, i.e. after the end of the transition period.

\section{Results and discussion}

\subsection{Tunnel temperature}

\begin{figure}[h]
	\centering
	\includegraphics{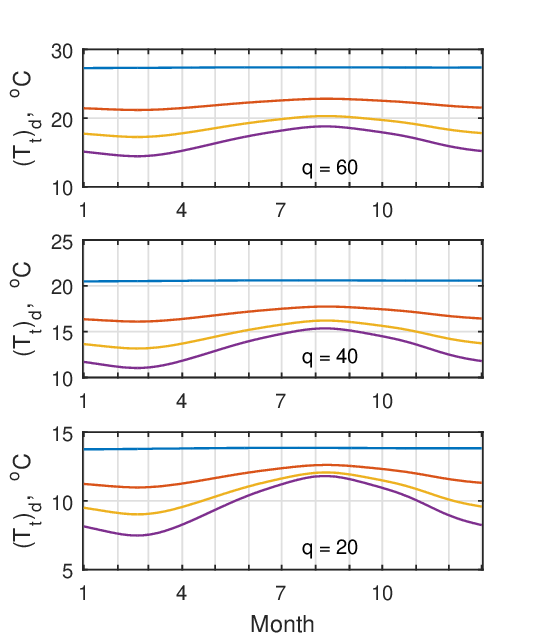}
	\caption{The average daily tunnel temperature $(T_t)_d$ at different values $q$ and $v$. In each panel $v$ = 0, 0.1, 0.2, 0.3 h$^{-1}$ -- curves from top to bottom.}
	\label{fig_tta}
\end{figure}

Heat generation and ventilation generally affect the air temperature in the tunnel. Heat generation tends to raise the air temperature and balance it with the ground temperature. Ventilation lowers the average temperature of the air in the tunnel and gives it the shape of a seasonal change in the temperature of the ambient air $T_a(t)$.  Both of these trends are clearly visible in the results of the calculations presented in Figure \ref{fig_tta}.

Figure \ref{fig_tta} shows graphs of the average daily air temperature in the tunnel in the "equilibrium" mode, obtained from the solution of the model for various values of the parameters $q$ and $v$. The calculations were carried out for three values of the power of heat sources $q$ = 20, 40, and 60 W/m, and four values of the ventilation rate $v$ = 0.0, 0.1, 0.2 and 0.3 h$^{-1}$.  Each panel in Fig. \ref{fig_tta} shows the results for one value of $q$.

At each value of parameter $q$, the maximum average air temperature in the tunnel is set in the absence of ventilation ($v = 0$ -- the uppermost curve on each panel).  With an increase in the ventilation rate $v$, the average temperature in the tunnel decreases and the temperature modulation in the tunnel by the temperature $T_a$ is more and more manifested. However, this modulation weakens with increasing $q$. 

The duration of the transition period also depends on the parameters $q$ and $v$ in different ways. It grows with increasing $q$ and decreases with increasing $v$. In particular, in the presented results, the largest transition period at 23 years is observed at $q=60$ W/m, $v = 0$, and the smallest at 6 years at $q=20$ W/m, $v = 0.3$ h$^{-1}$. Notice that such values of the transition period agree with duration of  "Dynamic Expansion Stage od Soil Heat Reservoir", reported in \cite{15}.

\subsection{Soil temperature}

Despite the fact that in our examples we consider the soil to be homogeneous, the established dynamically equilibrium temperature distribution in the soil is not obtained isotropic. The reason for this is the boundary conditions of the model, of which the presence of the second tunnel and heat exchange through the earth's surface have the main influence on the distortion of the temperature distribution. We will show some details of the temperature distribution in the soil using a calculation example for $q = 60$ W/m, $v = 0.2$ h$^{-1}$. In this case, the difference in average daily  temeratures in the tunnel during the year is about three degrees (figure \ref{fig_tta}), while in \cite{15}, this difference is about 17 degrees. As it follows from figure \ref{fig_tta}, the larger difference of average daily temperatures may be the result of larger ventilation rate and smaller power of heat generation in the tunnel.

\begin{figure}[hp]
	\centering
	\includegraphics{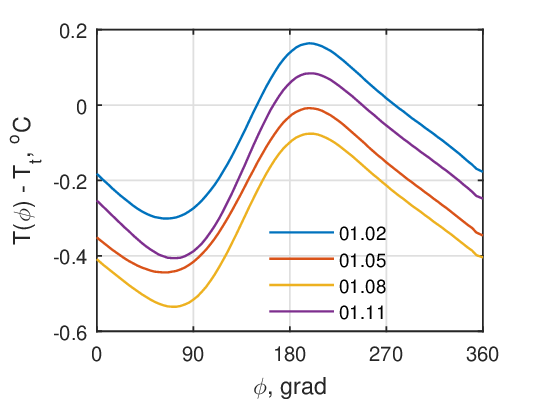}
	\caption{The temperature difference between the soil adjacent to the tunnel wall and the air temperature in the tunnel depending on the direction (angle $\phi$) for four dates in different seasons.}
	\label{fig_ttw}
\end{figure}

From the side of the tunnel, the anisotropy of the temperature distribution in the soil is expressed as a function of the angle $\phi$ (see Eq. (\ref{tav})). (The angle $\phi$ is counted counterclockwise from the positive direction of the axis $x$.) Figure 6 shows graphs of the temperature difference between the adjacent soil and the air temperature in the tunnel depending on the angle $\phi$ for four dates in different climatic seasons. It can be seen that in all seasons the maximum temperature difference is observed on the side facing the parallel tunnel, and the minimum - in the opposite direction. In the autumn and winter months, the temperature of the ground from the side of the adjacent tunnel becomes higher than the temperature of the tunnel and the heat goes from the soil to the tunnel, while on the rest of the wall of the tunnel, heat goes in the opposite direction - from the tunnel to the soil. Note that this situation is possible only with a certain ratio between the parameters $q$ and $v$. If, for example, in this case, the ventilation rate is gradually reduced, the temperature difference will always become negative, with the heat flux from the tunnel to the soil.
\begin{figure}[h]
	\centering
	\includegraphics{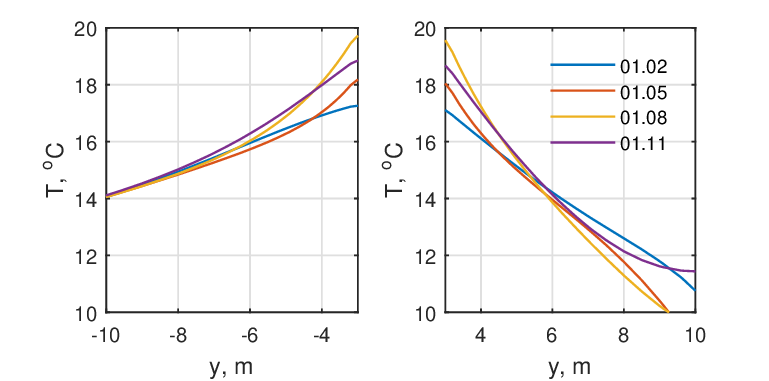}
	\caption{The surrounding soil temperature along $y$ axis in different seasons.}
	\label{fig_tsv}
\end{figure}

\begin{figure}[h]
	\centering
	\includegraphics{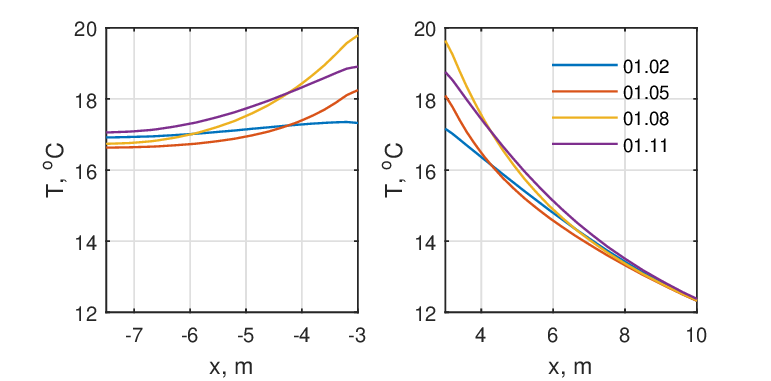}
	\caption{The surrounding soil temperature along $x$ axis in different seasons.}
	\label{fig_tsh}
\end{figure}

Figures \ref{fig_tsv} and \ref{fig_tsh} give an idea of the temperature distribution in the soil surrounding the tunnel.  Temperature plots along the vertical ($y$) axis are shown in Figure 7. Since a shallow tunnel is considered, the upward direction shows a significant influence of seasonal temperature changes at the earth's surface.  Nevertheless, at about $y=6$ (which corresponds to 3 meters above the tunnel and 9 meters below the earth's surface) the seasonal difference of the temperature is the smallest and  does not exceed 0.2 degrees.

In the downward direction, the temperature gradually decreases to the natural soil temperature regardless of the season, and at about $y=-10$ (7 meters under the tunnel at a depth of about 25 meters) the seasonal temperature difference becomes negligible.

In the horizontal direction (along the $x$-axis, Figure \ref{fig_tsh}), the temperature distribution in the soil is also asymmetric. The presence of a parallel tunnel plays a significant role here. Towards the parallel tunnel, the temperature drops much more slowly with increasing distance, and halfway between the tunnels, the ground temperature is practically constant throughout the year. Thus, a sufficiently stable high temperature "core" of the heat reservoir is formed in the area between the tunnels. The seasonal difference in the soil temperature becomes negligible at about $y=10$, which corresponds to 7 meters  from the tunnel wall in the horizontal direction opposite the central plane between the tunnels.

\begin{figure}[h]
	\centering
	\includegraphics{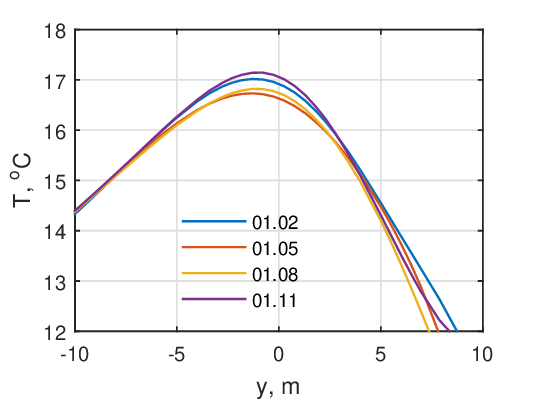}
	\caption{Soil temperature along the vertical centerline between tunnels.}
	\label{fig_tsm}
\end{figure}

According to the definition of the heat reservoir in \cite{15} as the maximal depth of the soil temperature changing, we can estimate that the heat reservoir for the selected values of the problem parameters is located at a depth of 9 to 25 meters below the earth's surface, and the width of the heat reservoir is about 20 meters. The heat reservoir surrounds the tunnels in such a way that its  boundary is about 3 meters above and 7 meters below the walls of the tunnels, and  is  at a distance of about 7 meters  from the walls of the tunnels in the horizontal direction. Note that the obtained dimensions of the heat reservoir are approximately 3 times less than the dimensions of the heat reservoir  in \cite{15}. Obviously, this is the result of a smaller annual temperature difference in the tunnel and anizopropy of the temperature field of the surrounding soil due to the influence of the earth's suface temperature and the presence of the second tunnel in our problem statement.

The vertical size of this reservoir along the cenral line between the tunnels can be estimated using Figure 9, which shows graphs of soil temperature along a vertical line through the midpoint of the tunnel spacing ($x=-D_t/2$). This means that the upper boundary of the heat reservoir crosses the centerline at about $y=4$   (which corresponds to 11 meters below the earth's surface) and is located lower than the bondary above the tunnels.

\section{Conclusion}

In conclusion, we proposed and analyzed a mathematical model for predicting the average daily temperature in the metro tunnel and the surrounding soil over a long period, which can reach several decades. In our model, we focused on three main factors that, in our opinion, play an important role in the formation of the thermal regime in the metro tunnel:
\begin{itemize}
\item{} heat generation,
\item{} heat exchange with adjacent soil, and
\item{} mechanical ventilation. 
\end{itemize}

The finite element method used to solve our mathematical model allows us to take into account the geological features of the soil by creating an adaptive spatial grid and using variable soil parameters. Time-dependent heat production and ventilation rates allow the model to take into account daily and seasonal cycles, as well as trends in metro operation and development.

The main simplifying assumption of the model, which allows the temperature to be calculated over long time intervals, is the assumption of ideal air mixing in the tunnel. This assumption allows us to avoid difficult calculations of air flows in the tunnel and to treat ventilation as a simple replacement of air in the tunnel with atmospheric air. In addition, we do not take into account the effect of moisture on the thermal properties of air and soil, however, taking into account moisture is not a fundamental difficulty and, as we believe, will not qualitatively change the results of calculations.

Numerical calculations presented in the work demonstrate the influence of various factors on the average temperature in the tunnel and in the surrounding soil. It is shown that the steady-state average annual temperature of the tunnel is directly proportional to the power of heat sources and inversely proportional to the ventilation rate. We also demonstrate the applicability of proposed model to  estimate the position and dimensions of the heat reservoir \cite{15}\, formed due to the operation of the underground transportation tunnel.

The results of calculations using the proposed model, taking into account specific circumstances, can be useful both in the design of new metro lines and in the development of existing lines or correction of their operating mode.

\section{Disclosure of Potential Conflicts of Interest}

The Authors declare that there is no conflict of interest.

\end{document}